\documentclass[pra,preprint,tightenlines,a4paper,showpacs]{revtex4}
\usepackage{bm,dcolumn,graphicx}
\begin{document}
\title{Accurate relativistic many-body calculations of van der Waals
coefficients $C_8$ and $C_{10}$ for alkali-metal dimers}
\author{Sergey G. Porsev}
\altaffiliation{Permanent Address: Petersburg Nuclear Physics
Institute, Gatchina, Leningrad district, 188300, Russia.}
\affiliation{Physics Department, University of Nevada, Reno,
Nevada 89557-0058.}
\author{Andrei Derevianko}
\affiliation{Physics Department, University of Nevada, Reno,
Nevada 89557-0058.}

\date{\today}

\begin{abstract}
We consider long-range interactions between two alkali-metal atoms
in their respective ground states. We extend the previous
relativistic many-body calculations of $C_6$ dispersion
coefficients [Phys.\ Rev.\ Lett.\ {\bf 82}, 3589 (1999)] to
higher-multipole coefficients $C_8$ and $C_{10}$. A special
attention is paid to usually omitted contribution of core-excited
states. We calculate this contribution within relativistic
random-phase approximation and demonstrate that for heavy atoms
core excitations contribute as much as 10\% to the dispersion
coefficients. We tabulate results for both homonuclear and
heteronuclear dimers and estimate theoretical uncertainties. The
estimated uncertainties for $C_8$ coefficients range from 0.5\%
for Li$_2$ to 4\% for Cs$_2$.
\end{abstract}

\pacs{34.20.Cf, 32.10.Dk, 31.15.Md, 31.15.Ar}

\maketitle

\section{Introduction}

We carry out accurate relativistic many-body atomic-structure
calculations of van der Waals interactions~\cite{DalDav66} between
alkali-metal atoms in their respective ground states. These
long-range interactions may be parameterized using dispersion (van
der Waals) coefficients $C_n$
\begin{eqnarray}
V(R) \approx -\frac{C_6}{R^6} - \frac{C_8}{R^8} -\frac{C_{10}}{R^{10}} + \cdots \, ,
\end{eqnarray}
where $R$ is the internuclear separation. A renewed interest in
high-accuracy interatomic potentials has been stimulated  by
advances in studies of ultracold collisions~\cite{WeiBagZil99}. At
low energies, collision properties are typically very sensitive to
details of the potentials. Thus accurate potentials are essential
for reliable {\em ab initio} description of ultracold collision
properties and, conversely, a wealth of information about the
potentials may be inferred from photoassociation and
Feshbach-resonance spectroscopy with ultracold atomic samples. In
particular, only recently interpretation of experiments with
ultracold atoms allowed several groups to reduce uncertainties in
the $C_6$ coefficients to a fraction of a per
cent~\cite{RobBurCla99,LeoWilJul00,AmiDulGut02}. These inferred
coefficients are in an excellent agreement with our values
predicted using many-body perturbation theory~\cite{DerJohSaf99}.
Even more refined understanding of details of ultracold collisions
led very recently to constraints on higher-multipole coefficient
$C_8$ for Rb~\cite{KemKokHei02,MarVolSch02} and
Cs~\cite{NISTprivate03}. This latest progress and discrepancies
between previous
determinations~\cite{MaeKut79,StaCer85,BusAub85,MarSadDal94,PatTan97}
of $C_8$ and $C_{10}$ coefficients motivate us to calculate these
coefficients using accurate relativistic many-body techniques of
atomic structure. In particular, we demonstrate that usually
omitted  contribution of core-excited states increases $C_n$ for
heavy atoms by as much as 10\%.

The main result of the paper --- compilation of van der Waals
coefficients $C_8$ and $C_{10}$  for homonuclear and heteronuclear
Li, Na, K, Rb, and Cs dimers is presented in
Tables~\ref{C8hom}--\ref{C10het}. The rest of the paper is
organized as follows. In Section \ref{SecFormalism} we present the
formalism. Numerical evaluation is discussed in
Section~\ref{SecDetails}. A detailed analysis of underlying
multipole dynamic and static polarizabilities is presented in
Section~\ref{SecPolar}. Finally, in Section~\ref{SecvdW} we
compile dispersion coefficients and estimate theoretical
uncertainties. Atomic units ($|e|=m_e=\hbar\equiv 1$) are used
throughout the paper.

\section{General formalism}
\label{SecFormalism}
The long-range part of electrostatic
interaction between two atoms $a$ and $b$ in their respective spherically-symmetric
states may be represented as~\cite{DalDav66}
\begin{equation}
V(R)=-\sum_{n=3} C^{ab}_{2n}/R^{2n} \, ,                                    
\label{V_R}
\end{equation}
where $R$ is the distance between atoms. For ground-state atoms van der Waals
coefficients  are given  by~\cite{PatTan00}
\begin{equation}
C^{ab}_{2n}= \frac{(2n-2)!}{2\pi}\,
\sum_{l=1}^{n-2}\,\frac{1}{(2l)!\,(2l')!}
\int_0^\infty \alpha_l^a(i \omega)\, \alpha_{l'}^b(i \omega) d\omega \, ,   
\label{C2n}
\end{equation}
where $l'=n-l-1$; $\alpha_l^a(i \omega)$ and $\alpha_{l'}^b(i
\omega)$ are, respectively, $2^l$-pole dynamic polarizability of
atom $a$ and $2^{l'}$-pole dynamic polarizability of atom $b$. The
dynamic polarizabilities in Eq.~(\ref{C2n}) are defined as
\begin{equation}
\alpha_l ( i \omega ) = 2\,{\rm Re} \sum_k \frac
{\langle \Psi_0 |T^{(l)}_0 | \Psi_k \rangle
 \langle \Psi_k | T^{(l)}_0 | \Psi_0 \rangle}                               
  {\left( E_k-E_0 \right) + i \omega } \, .
\label{alpha1}
\end{equation}
Here the summation extends over a complete set of atomic states and
$T^{(l)}_0$ are the zeroth components of spherical tensors of electric-multipole operators
\begin{equation}
T^{(l)}_m = \sum_{i=1}^N r_i^l C^{(l)}_m\left( \hat{\mathbf{r}}_i \right) \, ,
\end{equation}
where $C^{(l)}_m$ are normalized spherical
harmonics~\cite{VarMosKhe88} and the sum is over  all $N$ atomic
electrons.

Previously many-body calculations of dispersion coefficients $C_6$
were carried out in Refs.~\cite{DerJohSaf99,DerBabDal01}, and here
we focus on dispersion coefficients $C_8$ and $C_{10}$. As follows
from an examination of Eq.~(\ref{C2n}), we need to compute dipole
$\alpha_1$, quadrupole $\alpha_2$, and octupole $\alpha_3$ dynamic
polarizabilities. In this work we employ dynamic dipole
polarizabilities calculated previously in Ref.~\cite{DerJohSaf99}
and determine higher-multipole polarizabilities $\alpha_2$ and
$\alpha_3$.

Following~\cite{DerJohSaf99}
we separate all intermediate states in the sum Eq.~(\ref{alpha1})
into valence and core-excited states
\begin{equation}
\alpha_l ( i \omega ) = \alpha_l^v ( i \omega ) + \alpha_l^c ( i \omega)
                      + \alpha_l^{cv} ( i \omega ) \, ,
\label{alphaBreak}
\end{equation}
Here $\alpha_l^v ( i \omega)$ is a traditional term encapsulating
excitations of the valence electron. Contributions of
electric-multipole excitations of core electrons are denoted by
$\alpha_l^c ( i \omega)$. Finally, a small counter term
$\alpha_l^{cv} ( i \omega )$ is related to excitations of core
electrons to occupied valence state. We include these
exclusion-principle-forbidden excitations in the calculations of
core polarizabilities and thus we have to introduce the counter
term (see Ref.~\cite{DerPor02Cs} for more details). We will
discuss calculations of the $\alpha_l^v ( i \omega)$  and
$\alpha_l^c ( i \omega)$ terms later on. Here we just briefly
comment on the counter term  $\alpha_l^{cv} ( i \omega )$. For
octupole polarizabilities $\alpha_3^{cv} ( i \omega )$ term simply
vanishes in independent-particle approximation since E3 selection
rules would require an excitation from $f$ shell to valence
$s$-state and none of the alkalis considered here (Li through Cs)
has filled f-shells. Since we employ dipole polarizabilities from
Ref.~\cite{DerJohSaf99}, the counter term, calculated in
Dirac-Hartree-Fock (DHF) approximation is included in $\alpha_1 (
i \omega )$. Finally we disregard this correction for quadrupole
polarizabilities, it gives a negligible contribution due to
required excitation of deeply bound $d$ electrons from the core.

High-accuracy calculations of the {\em dipole} dynamic
polarizabilities were carried out earlier in
Ref.~\cite{DerJohSaf99} and we employ these dipole
polarizabilities in the present work. In those calculations a
combination of several relativistic many-body techniques was
employed. A dominant contribution to $\alpha_1^v$ has been
calculated with all-order linearized coupled-cluster method
truncated at single and double excitations. High-accuracy
experimental values for energies and electric-dipole matrix
elements for principle transitions has been employed to refine the
dipole polarizabilities. In the following we focus on the
quadrupole and octupole polarizabilities.

To find the quadrupole $\alpha_2^v$ and octupole $\alpha_3^v$
valence contributions we applied a relativistic many-body method
initially suggested in Refs.~\cite{DzuFlaKoz96a,DzuFlaKoz96b} and
subsequently developed
in~\cite{DzuKozPor98,PorRakKoz99J,PorRakKoz99P,KozPor99E}. In this
method one determines wave functions from solution of the
effective many-body Shr\"{o}dinger equation
\begin{equation}
H_{\rm eff}(E_n) \, | \Psi_n \rangle = E_n \, |\Psi_n \rangle \, ,
\label{H}
\end{equation}
with the effective Hamiltonian defined as
\begin{equation}
  H_{\rm eff}(E) = H_\mathrm{FC} + \Sigma(E) \, .
\label{Heff}
\end{equation}
Here $H_\mathrm{FC}$ is the frozen-core Dirac-Hartree-Fock
Hamiltonian and self-energy operator $\Sigma$ is the
energy-dependent correction, involving core excitations.
Qualitatively $\Sigma$ operator corresponds to  core polarization
term in model potentials employed in
Refs.~\cite{MarSadDal94,PatTan97}. In the present calculation the
self-energy operator recovers second order of perturbation theory
in residual Coulomb interaction and additionally accounts for
certain classes of many-body diagrams in all orders of
perturbation theory.

The concept of effective Hamiltonian $H_{\rm eff}$ may be extended
to other operators. We introduce effective  (or dressed)
electric-multipole operators $T^l_{\rm eff}$ acting in the model
space of valence electrons.  These operators were obtained within
the  relativistic random-phase approximation
(RRPA)~\cite{DzuKozPor98,KolJohSho82,JohKolHua83}. Qualitatively,
the RRPA describes a shielding of the externally applied
electric-multipole field by the core electrons. The RRPA sequence
of diagrams was summed to all orders of the perturbation theory.

Once the ground-state wavefunctions are obtained from
Eq.~(\ref{H}), the dynamic {\em valence} polarizabilities
$\alpha_l^v(i \omega)$ are computed with the
Sternheimer~\cite{Ste50} or Dalgarno-Lewis \cite{DalLew55} method
implemented in  the DHF+$\Sigma$+RRPA framework. (In the following
we denote $\Sigma$+RRPA corrections as the many-body perturbation
theory (MBPT) corrections.) Given ground-state wave-function
$\Psi_0$ and energy $E_0$, we find an intermediate-state wave
function $\Psi_f$ from an inhomogeneous equation
\begin{eqnarray}
|\Psi_f \rangle & = &
{\rm Re}\, \left\{ \frac{1}{H_{\rm eff} -E_0+i \omega}\,
 \sum_i | \Psi_i \rangle
 \langle \Psi_i | (T^l_0)_{\rm eff} |\Psi_0 \rangle \right\} \nonumber \\
&=&
{\rm Re}\, \left\{ \frac{1}{H_{\rm eff}-E_0+i \omega}\,
                  (T^l_0)_{\rm eff} |\Psi_0 \rangle \right\}.
\label{psif}
\end{eqnarray}
With such introduced $\Psi_f$  Eq.~(\ref{alpha1}) becomes simply
\begin{equation}
\alpha_l^v ( i \omega ) = 2\, \langle \Psi_0 |(T^l_0)_{\rm eff}| \Psi_f \rangle \, ,
\label{alpha2}
\end{equation}
where subscript $v$ emphasized that only excitations of the
valence electron to higher virtual orbitals are included in the
intermediate-state wave function $\Psi_f$ due to a presence of
$H_{\rm eff}$  in Eq.~(\ref{psif}). As to additional contribution
$\alpha^c_l$ of core-excited states, we employ the relativistic
random-phase approximation method described in
Refs.~\cite{KolJohSho82,JohKolHua83}.

\section{Details of numerical calculation}
\label{SecDetails} At the first stage of calculations we
determined core orbitals and valence orbitals for several
low-lying states from the frozen-core Dirac-Hartree-Fock
equations~\cite{BraDeyTup77}. The virtual orbitals were determined
with the help of a recurrent procedure~\cite{PorKozRak01}.
One-electron basis sets of the following sizes were used on the
stage DHF+$\Sigma$ calculations:
$$ {\rm Li}: \quad 1-17s, \, 2-17p, \, 3-16d, \, 4-16f, \, 5-10g;$$
$$ {\rm Na}: \quad 1-18s, \, 2-18p, \, 3-17d, \, 4-17f, \, 5-11g;$$
$$  {\rm K}: \quad 1-19s, \, 2-19p, \, 3-18d, \, 4-19f, \, 5-12g;$$
$$ {\rm Rb}: \quad 1-20s, \, 2-20p, \, 3-19d, \, 4-19f, \, 5-13g;$$
$$ {\rm Cs}: \quad 1-23s, \, 2-23p, \, 3-23d, \, 4-26f, \, 5-14g.$$

Using these basis sets we solved the multi-particle Shr\"{o}dinger
equation (\ref{H}) and found the wave functions of low-lying
states. As discussed in~\cite{KozPor99O} and demonstrated
in~\cite{PorKozRak00Z,PorKozRak01} a proper approximation for the
effective Hamiltonian can substantially improve an agreement
between calculated and experimental spectra of multielectron atom.
One can introduce an energy shift $\delta$ and replace $\Sigma(E)
\rightarrow \Sigma(E-\delta)$ in the effective Hamiltonian,
Eq.~(\ref{Heff}). We have determined $\delta$ from a fit of
theoretical energy levels to experimental spectrum. Using only one
fitting parameter $\delta$ we reproduced the experimental energies
for 12 low-lying states for Li and for 10 low-lying states for Na
and K with accuracy 0.1--0.2\%. To reproduce the low-lying energy
levels with the same 0.1-0.2\% accuracy for heavier Rb and Cs we
used three fitting parameters (different shifts $\delta$ for
different partial waves). An illustrative comparison for the
heaviest atom Cs (55 electrons) is presented in Table~\ref{Cs_E}.
It is worth noting that an empirical introduction of shifts
$\delta$ mimics higher-order many-body corrections in perturbation
theory. We will estimate theoretical uncertainty based on
sensitivity of our results to variation in these shifts.
\begin{table}
\caption{Comparison of DHF and many-body one-electron removal
energies $E_\mathrm{val}$ for Cs with experimental values.
$E_\mathrm{val}$ are given in atomic units. $\Delta$ are
excitation energies from the ground $6s_{1/2}$ state in
$\mathrm{cm}^{-1}$. For $s$-states the energies were calculated
with $\delta= -0.20$ a.u., for $d$-states - with $\delta= 0.0$
a.u., and for $p$   with $\delta= -0.09$ a.u..}
\label{Cs_E}
\begin{ruledtabular}
\begin{tabular}{ldcdcdc}
& \multicolumn{2}{c}{\qquad DHF} & \multicolumn{2}{c}{\qquad DHF+MBPT}
& \multicolumn{2}{c}{\qquad Experiment~\cite{Moo71}} \\
Config.&\multicolumn{1}{c}{\qquad E$_{\rm val}$} &\multicolumn{1}{c}{$\Delta$}
       &\multicolumn{1}{c}{\qquad E$_{\rm val}$} &\multicolumn{1}{c}{$\Delta$}
       &\multicolumn{1}{c}{\qquad E$_{\rm val}$} &\multicolumn{1}{c}{$\Delta$}  \\
\hline
$6s_{1/2}$  &  0.127368  &   ---     & 0.143085 &   ---     &  0.143099\footnotemark[1]& --- \\
$6p_{1/2}$  &  0.085616  &   9163.6  & 0.092173 &  11172.2  &  0.092167  & 11178.2 \\
$6p_{3/2}$  &  0.083785  &   9565.3  & 0.089609 &  11734.9  &  0.089642  & 11732.4 \\
$5d_{3/2}$  &  0.064419  &  13815.7  & 0.076995 &  14503.3  &  0.077035  & 14499.5 \\
$5d_{5/2}$  &  0.064529  &  13791.5  & 0.076459 &  14621.0  &  0.076590  & 14597.1 \\
$7s_{1/2}$  &  0.055187  &  15841.8  & 0.058475 &  18568.0  &  0.058645  & 18535.5 \\
$7p_{1/2}$  &  0.042021  &  18731.4  & 0.043868 &  21773.9  &  0.043928  & 21765.7 \\
$7p_{3/2}$  &  0.041368  &  18874.8  & 0.043041 &  21955.4  &  0.043103  & 21946.7 \\
\end{tabular}
\end{ruledtabular}
\small \footnotemark[1]{For~the~ground state
E$_{\rm val}$ = IP~(Cs) = 31406.71 cm$^{-1}$ \cite{Moo71}}.
\end{table}


\section{Quadrupole and octupole polarizabilities}
\label{SecPolar}

To reiterate major steps of the formalism described in
Section~\ref{SecFormalism}, we determined ground state wave
functions  from the effective many-body Shr\"{o}dinger
equation~(\ref{H}), calculated dressed electric multipole
operators $T_\mathrm{eff}$, solved inhomogeneous equation
(\ref{psif}) and computed valence parts $\alpha_l^v$ of dynamic
polarizability with Eq.~(\ref{alpha2}). Additional contributions
$\alpha_l^c$ of core-exited states were calculated using RRPA
method.

Calculation of dynamic polarizabilities with $\omega=0$ gives us
the static polarizabilities. We provide these data in
Tables~\ref{alpha_stat_q} and \ref{alpha_stat_o} and compare them
with other results. To estimate uncertainties we present in the
Tables results of pure DHF calculations and compare them with
DHF+MBPT ones. The uncertainties of calculations are associated
with higher orders of the MBPT which are taken into account only
partially. The heavier the atom, the larger MBPT contribution is and
we expect theoretical accuracy to become worse. For instance, the MBPT
correction to the static quadrupole polarizability $\alpha_2^v$
for Li is only 4\%, while for Cs it attains 38\%. For static
octupole polarizabilities $\alpha_3^v$ the MBPT corrections are
larger and range from 5\% for Li to 48\% for Cs.

Let us turn to estimates of theoretical uncertainty of quadrupole
polarizabilities. Essentially it is  based on sensitivity of our
results to semiemprically introduced shifts $\delta$. As mentioned
in Section~\ref{SecDetails} an introduction of these shifts mimics
omitted higher-orders of perturbation theory. We estimate the
theoretical error bar as a half of the difference between {\em ab
initio} ($\delta=0$) value and result with semiempirically chosen
$\delta$. Further an overwhelming contribution to static
$2^l$-pole polarizabilities Eq.(\ref{alpha1}) comes from the
lowest-lying valence state of proper angular symmetry. Since we
recover experimental energies almost exactly (see
Table~\ref{Cs_E}), the theoretical uncertainty is  determined by
an accuracy of calculation for electric-multipole operators of
principal transitions. We write
\[
 \frac{\delta \alpha_2(0)}{\alpha_2(0)} \sim
 \frac{ \langle ns | T^2_0| n'd \rangle_{\delta} -
 \langle ns | T^2_0| n'd \rangle_{\delta=0}}{ \langle ns | T^2_0| n'd \rangle_{\delta=0}},
\]
where $ns$ denotes the ground state and  $n'd$ stands for
lowest-lying valence d-states. For example, following this
procedure we obtain an error bar of 0.3\% for Li. Our result of
1424(4) for Li is in excellent agreement with the value
1423.266(5)  from  benchmark high-accuracy variational
non-relativistic calculations by \citet{YanBabDal96}. We estimate
theoretical uncertainties for octupole polarizabilities to be at
10\% level for heavy atoms.
\begin{table}
\caption{Static quadrupole polarizabilities $\alpha_2$ for ground states
of alkali-metal atoms in a.u. We present valence contributions for the
cases of pure DHF and DHF+MBPT, and core contributions. Final values
were determined as sum of $\alpha_2^v$ (DHF+MBPT) and $\alpha_2^c$.}
\label{alpha_stat_q}
\begin{ruledtabular}
\begin{tabular}{lccccc}
 & \multicolumn{1}{c}{Li}  & \multicolumn{1}{c}{Na}
 & \multicolumn{1}{c}{K }  & \multicolumn{1}{c}{Rb}
 & \multicolumn{1}{c}{Cs}  \\
\hline
{$\alpha_2^v$ (DHF)}                              & 1485.5 & 2230.3 & 7049 & 9790 & 16613  \\
{$\alpha_2^v$ (DHF+MBPT)}                         & 1424.5 & 1883.6 & 4983 & 6488 & 10388  \\
{$\alpha_2^c$ (RRPA)}                             &    0.1 &    1.5 &   16 &   35 &    86  \\
{Final}                                & 1424(4) & 1885(26) & 5000(45) & 6520(80) & 10470(390) \\
                     \multicolumn{6}{c}{Other works} \\
Patil and Tang~\protect\cite{PatTan99}            & 1393   & 1796   & 4703 & 6068 & 10260  \\
Patil and Tang~\protect\cite{PatTan97}            & 1403   & 1807   & 4760 & 6163 & 10400  \\
Yan {\it et al.}~\protect\cite{YanBabDal96}       & 1423.266(5)&    &      &      &        \\
Marinescu {\it et al.}~\protect\cite{MarSadDal94} & 1424   & 1878   & 5000 & 6495 & 10462  \\
Spelsberg {\it et al.}~\protect\cite{SpeLorMey93} & 1423   & 1879   & 5001 & ---  &   ---   \\
Maeder and Kutzelnigg~\protect\cite{MaeKut79}     & 1383   & 1799   & 4597 & 5979 & 9478    \\
\end{tabular}
\end{ruledtabular}
\end{table}
\begin{table}
\caption{Static octupole polarizabilities $\alpha_3$ for ground
states of alkali-metal atoms in $10^4$ a.u. We present valence
contributions for the cases of pure DHF and DHF+MBPT, and core
contributions. Final values were determined as sum of $\alpha_3^v$
(DHF+MBPT) and $\alpha_3^c$.}
\label{alpha_stat_o}
\begin{ruledtabular}
\begin{tabular}{lddddd}
 & \multicolumn{1}{c}{Li}  & \multicolumn{1}{c}{Na}
 & \multicolumn{1}{c}{K }  & \multicolumn{1}{c}{Rb}
 & \multicolumn{1}{c}{Cs}  \\
\hline
{$\alpha_3^v$ (DHF)}         &   4.185   &  6.888   &   28.10   &  41.50    & 76.49    \\
{$\alpha_3^v$ (DHF+MBPT)}    &   3.957   &  5.536   &   17.73   &  23.66    & 39.43    \\
{$\alpha_3^c$ (RRPA)}        &      0    &  0.001   &    0.01   &   0.03    &  0.10    \\
{Final}                      & 3.957 & 5.54 & 17.7 & 23.7 & 39.5 \\
                     \multicolumn{6}{c}{Other works} \\
\protect\citet{PatTan99}     & 3.871     & 5.287    & 16.07     & 20.73     & 33.12     \\
\protect\citet{PatTan97}     & 3.986     & 5.430    & 16.30     & 20.97     & 33.33     \\
\protect\citet{YanBabDal96}  & 3.965049(8)&         &           &           &           \\
\protect\citet{MarSadDal94}  & 3.969     & 5.552    & 17.69     & 23.69     & 39.53     \\
\protect\citet{SpeLorMey93}  & 3.927     & 5.486    & 19.14     &           &            \\
\protect\citet{MaeKut79}     & 3.680     & 5.117    & 15.02     & 21.27     & 33.99      \\
\end{tabular}
\end{ruledtabular}
\end{table}
Our results for static polarizabilities are listed in
Tables~\ref{alpha_stat_q} and \ref{alpha_stat_o}. In these Tables
we also compare our results with the predictions by other authors.
We find that for light atoms there is a good agreement between
different results except the values obtained by~\citet{MaeKut79}
are consistently smaller. As the number of atomic electrons
increases, the correlation effects become more pronounced and
discrepancies between results from different groups grow larger.
\citet{MarSadDal94} used a model potential with five adjustment
parameters obtained by fitting to experimental energy levels.
Core-polarization was included in the pseudo-potential and they
also included effects of shielding (or field dressing) in the
multipole operators. \citet{PatTan97} also used effective
potential in their calculations to obtain the wave functions of
excited states, but they used one-parametric potential and did not
shielding in the multipole operators. Generally, our results are
in a good agreement with all results except for values
by~\citet{MaeKut79}. One of possible reasons for this discrepancy
is that these authors used very small number of basis functions
(e.g. only 5 basis orbitals for $p$, $d$, and $f$ partial waves)
while $\alpha_2$ and $\alpha_3$ polarizabilities are very
sensitive to details of construction and saturation of basis sets.

Also shown in Tables~\ref{alpha_stat_q} and \ref{alpha_stat_o} are
the corrections $\alpha_l^c$ due to core-excited states. These
quantities are essentially  polarizabilities of singly-charged
ions of alkali-metal atoms. Only disregarding distortion of the
core by  the valence electrons, one may identify corrections
$\alpha_l^c$ as core polarizabilities. For static quadrupole
polarizabilities their relative contribution to the total
polarizabilities ranges from 0.01\% for Li to 0.8\% for Cs. The
core corrections to static octupole polarizabilities are even
smaller (just 0.25\% for Cs). Relative smallness of $\alpha_l^c$
terms for {\em static} polarizabilities may lead one to a wrong
assumption that the core excitations may be  disregarded  in
calculations of van der Waals coefficients $C_n$. In fact the
expression (\ref{C2n}) for $C_n$ contains integration over an
infinite range of frequencies $\omega$. While the region around
$\omega=0$ does provide the dominant contribution to $C_n$, the
high-frequency tail of the polarizability is still important. As
$\omega \rightarrow \infty$ the core polarizability overpowers
valence contribution. In fact, one of the points of the
paper~\cite{DerJohSaf99} was to explicitly demonstrate that for
heavy atoms the core polarizability may contribute as much as 15\%
to $C_6$ dispersion coefficient. Here  using RRPA calculations of
$\alpha_l^c(i \omega)$ core polarizability we will arrive at a
similar conclusion for higher-multipole coefficients $C_8$ and
$C_{10}$.

We calculated the core polarizabilities in the framework of
relativistic random-phase approximation method (RRPA). Essentially
we extended approach of \citet{JohKolHua83} and incorporated
frequency dependence into the calculations. Compared to
Ref.~\cite{JohKolHua83} we also employed a different numerical
technique using B-spline basis sets. With our newly developed code
we recover the previous results~\cite{JohKolHua83} for static
dipole and quadrupole polarizabilities. We found that unusually
large basis sets of 100 B-splines were required to achieve a
numerical convergence, especially for octupole polarizabilities of
heavy atoms. Finally, we present a comparison of the computed RRPA
static quadrupole and octupole core polarizabilities with other
works in Tables~\ref{alpha2_c} and \ref{alpha3_c}.
Patil~\cite{Pat94,Pat86} has inferred these polarizabilities
analyzing Rydberg spectra of alkalis. His results are in a uniform
agreement with our {\em ab initio} values.
\begin{table}
\caption{Static quadrupole polarizabilities  $\alpha_2^c(0)$ of
singly-charged ions of alkali-metal atoms (core polarizabilities).
Results marked RRPA are results of our calculations; these
numerical values are identical to those by
\protect\citet{JohKolHua83}. All values are in atomic units.}
\label{alpha2_c}
\begin{ruledtabular}
\begin{tabular}{lcddccc}
 & \multicolumn{1}{c}{Li$^+$}  & \multicolumn{1}{c}{Na$^+$}
 & \multicolumn{1}{c}{K$^+$ }  & \multicolumn{1}{c}{Rb$^+$}
 & \multicolumn{1}{c}{Cs$^+$}  \\
\hline
{RRPA}                      & 0.11 &  1.52         &   16.3   & 35.4 & 86.4  \\
\protect\citet{Pat94,Pat86} &      &  1.64(15)    & 18.2(3.0)& 42(3)&128(40)\\
\protect\citet{FreKle76}    &      &  1.91(15)&          &      &       \\
\end{tabular}
\end{ruledtabular}
\end{table}
\begin{table}
\caption{Static octupole polarizabilities  $\alpha_3^c(0)$ of
singly-charged ions of alkali-metal atoms (core polarizabilities).
All values are in atomic units.}
\label{alpha3_c}
\begin{ruledtabular}
\begin{tabular}{lccclll}
 & \multicolumn{1}{c}{Li$^+$}  & \multicolumn{1}{c}{Na$^+$}
 & \multicolumn{1}{c}{K$^+$ }  & \multicolumn{1}{c}{Rb$^+$}
 & \multicolumn{1}{c}{Cs$^+$}  \\
\hline
{This work}                   & 0.17 & 7.5 & 110    &  314    & 1014     \\
Patil~\protect\cite{Pat94}    &      &     & 95(10) & 280(40) & 1220(200) \\
\end{tabular}
\end{ruledtabular}
\end{table}

\section{van der Waals coefficients }
\label{SecvdW}

{From} general formula~(\ref{C2n}) dispersion coefficients  may be
expressed as
\begin{eqnarray}
C^{ab}_6 &=& C_{ab}(1,1),    \nonumber \\
C^{ab}_8 &=& C_{ab}(1,2)+C_{ab}(2,1), \nonumber \\
C^{ab}_{10} &=& C_{ab}(2,2) + C_{ab}(1,3) + C_{ab}(3,1) \, .
\label{vdW}
\end{eqnarray}
Here the coefficients $C_{ab}(l,l')$ are
quadratures of atomic $2^{l}-$ and $2^{l'}-$pole dynamic polarizabilities
\begin{equation}
C_{ab}(1,1) = \frac{3}{\pi}\,
\int_0^\infty\, \alpha_1^a(i \omega)\, \alpha_1^b(i \omega) d\omega,
\label{C_11}
\end{equation}
\begin{equation}
C_{ab}(1,2) = \frac{15}{2\pi}\,
\int_0^\infty\, \alpha_1^a(i \omega)\, \alpha_2^b(i \omega) d\omega,
\label{C_12}
\end{equation}
\begin{equation}
C_{ab}(2,2) = \frac{35}{\pi}\,
\int_0^\infty\, \alpha_2^a(i \omega)\, \alpha_2^b(i \omega) d\omega,
\label{C_22}
\end{equation}
\begin{equation}
C_{ab}(1,3) = \frac{14}{\pi}\,
\int_0^\infty\, \alpha_1^a(i \omega)\, \alpha_3^b(i \omega) d\omega.
\label{C_13}
\end{equation}
Calculations of dynamic polarizabilities were discussed in the
previous section and here we proceed to evaluation of the
dispersion coefficients.

The computed $C_8$ and $C_{10}$ coefficients for homonuclear and
heteronuclear species are presented in Tables~\ref{C8hom}--
\ref{C10het}. The dispersion coefficients $C_6$ were tabulated
previously in Refs.~\cite{DerJohSaf99, DerBabDal01}. This
completes the first application of relativistic many-body methods
of atomic structure to calculations of leading long-range
interactions between ground-state alkali-metal atoms.

To  estimate uncertainties in our values we notice that the main
value of the quadratures, Eqs.~(\ref{C_11})--(\ref{C_13}) is
accumulated in the low-frequency region $\omega \approx 0$.
Therefore the error may be expressed via  uncertainties in the
static multipole polarizabilities
\[
 \frac{ \delta C_{ab}(l,l')}{C_{ab}(l,l')}  =
 \left\{  \left(\frac{ \delta \alpha_l(0)}{ \alpha_l(0) }\right)^2 +
  \left( \frac{ \delta \alpha_{l'}(0)}{ \alpha_{l'}(0) }\right)^2 \right\}^{1/2} \, .
\]
The required uncertainties $\delta \alpha_l(0)$ were estimated in Section~\ref{SecPolar}
and Ref.~\cite{DerJohSaf99}.
The error induced in $C_8^{ab}$ is
\[
\delta C^{ab}_8 = \left\{ (\delta C_{ab}(1,2))^2+  (\delta C_{ab}(2,1))^2
\right\}^{1/2} \, .
\]
Here we assumed that $a\ne b$. The formulas for homonuclear dimers
may be derived in a similar manner. The resulting theoretical
uncertainties for $C_8$ coefficients range  from 0.5\% for Li$_2$
to 4\% for Cs dimer. We anticipate uncertainty in $C_{10}$
coefficients to be better than 10\%.

It is instructive to consider the effect of core excitation
contribution $\alpha_l^c ( i \omega )$ to dynamic polarizabilities
and thus to $C_n$ coefficients. Such corrections are omitted in
the model potential calculations such as
Ref.~\cite{MarSadDal94,PatTan97}. To illuminate the relative
contributions of core-excitations we computed $C_n$ coefficients
by keeping only the valence contributions to the total dynamic
polarizabilities
$$ \alpha_l ( i \omega ) \rightarrow \alpha_l^v ( i \omega ). $$
Such calculated dispersion coefficients are marked as $C_8^v$ and
$C_{10}^v$ in Tables~\ref{C8hom}--\ref{C10het}, while values
marked ``final'' were obtained with an additional inclusion of
core excitations. Comparing these values, we observe that relative
contribution of $\alpha_l^c ( i \omega )$ term grows rapidly as
the number of atomic electrons increases. For example, examining
Table~\ref{C8hom} we see that core correction to $C_8$ for Li is
only 0.2\%, while for Cs it is 10\%. For $C_{10}$ coefficients the
core contributions for all atoms are slightly smaller. Still for
Cs  core excitations  contribute  8\% to the $C_{10}$ coefficient.

A comparison with results by other authors is presented in
Tables~\ref{C8hom}--\ref{C10het}. There is good agreement for
light Li and Na atoms. For heavier atoms, in particular for Cs,
there is discrepancy at the level of 10\% for $C_8$ and 20\% for
$C_{10}$ coefficients. Such tendency may be attributed to two
factors. First, correlations become enhanced for heavier atoms.
Another cause is that model-potential calculations such as
Ref.~\cite{MarSadDal94,PatTan97} disregard contribution of
core-excited states. This corresponds to the valence term denoted
as $C^v_n$ in Tables~\ref{C8hom}--\ref{C10het}. As mentioned above
the core-excited states contribute at the level of 10\% for Cs. If
we disregard this contribution, we see that the model-potential
results are in a reasonable agreement with our $C^v_n$ values.

Only very recently interpretation of experiments with ultracold
atoms allowed several groups to reduce uncertainties in the $C_6$
coefficients to a fraction of a per
cent~\cite{RobBurCla99,LeoWilJul00,AmiDulGut02}. These inferred
coefficients are in an excellent agreement with our values
predicted using many-body perturbation theory~\cite{DerJohSaf99}.
Even more refined understanding of details of ultracold collisions
led very recently to constraints on higher-multipole coefficient
$C_8$ for Rb$_2$~\cite{KemKokHei02,MarVolSch02} and Cs
dimer~\cite{NISTprivate03}. In Table~\ref{C8hom} we present a
comparision with these inferred values. Our computed value for
Rb$_2$ 5.77(8) agrees well with $C_8=5.79(49)$ by
~\citet{KemKokHei02} and $C_8=5.73$ by \citet{MarVolSch02}.
However, we disagree with 1\%-accurate result~\cite{KemKokHei02}
of 6.09(7) by four standard deviations. This 1\%-accurate result
was obtained in Ref.~\cite{KemKokHei02} by setting additional
constraints on the singlet potential of Rb dimer while including
higher-multipole van der Waals coefficients $C_{11}$ and $C_{12}$
in the fit. For Cs$_2$ the inferred value by~\citet{NISTprivate03}
is $C_8=8.4(4)$, it disagrees with our prediction, 10.2(4) by more
than four standard deviations. It is worth noting that while for
Rb the inferred value lies above our result, for Cs the situation
is reversed and our value is larger.

To conclude, we calculated static and dynamic quadrupole and
octupole polarizabilities for ground states of Li, Na, K, Rb, and
Cs atoms. The calculations were carried out using accurate
relativistic many-body methods of atomic structure. With the
computed polarizabilities we evaluated $C_8$ and $C_{10}$ van der
Waals coefficients for homonuclear and heteronuclear dimers and
estimated theoretical uncertainties. The estimated uncertainties
for $C_8$ coefficients range from 0.5\% for Li$_2$ to 4\% for
Cs$_2$. We have highlighted the role of usually omitted core
excitations in calculation of $C_8$, and $C_{10}$ coefficients and
found that their contribution is important for heavy atoms K, Rb,
and Cs.

We would to thank Mikhail Kozlov for comments on the manuscript.
This work was supported in part by the National Science Foundation.
The work of S.G.P. was additionally supported by the Russian
Foundation for Basic Research under Grant No. 02-02-16837-a.

\begin{table}
\caption{van der Waals $C_8$ coefficients in $10^5$ a.u. for
homonuclear dimers. $C_8^v$ values include only valence
contributions. The final values were determined as combination of
DHF+MBPT method for valence contributions with RRPA calculations
for core excitations.}
\label{C8hom}
\begin{ruledtabular}
\begin{tabular}{lcccccc}
 & \multicolumn{1}{c}{Li}  & \multicolumn{1}{c}{Na}
 & \multicolumn{1}{c}{K }  & \multicolumn{1}{c}{Rb}
 & \multicolumn{1}{c}{Cs}  \\
\hline
{$C_8^v$}                    & 0.832    & 1.15      & 4.00      &  5.37     & 9.16     \\
{Final}                      & 0.834(4) & 1.160(18) & 4.20(5)   &  5.77(8)  & 10.2(4) \\
                     \multicolumn{6}{c}{Other theoretical works}                      \\
\protect\citet{PatTan97}     & 0.8183   & 1.090   & 3.892     &  5.258    & 9.546     \\
\protect\citet{YanBabDal96}  & 0.834258(4)&       &           &           &           \\
\protect\citet{MarSadDal94}  & 0.8324   & 1.119   & 4.096     &  5.506    & 9.630     \\
\protect\citet{SpeLorMey93}  & 0.8303   & 1.141   & 4.011     &           &           \\
\protect\citet{MaeKut79}     & 0.8089   & 1.098   & 3.834     &  5.244    & 9.025     \\
                     \multicolumn{6}{c}{Experiment}                                   \\
\protect\citet{KemKokHei02}  &          &         &           &  5.79(49) &            \\
                             &          &         &           &  6.09(7)  &            \\
\protect\citet{MarVolSch02}  &          &         &           &  5.73     &            \\
\protect\citet{NISTprivate03} &          &         &           &           &  8.4(4)
\end{tabular}
\end{ruledtabular}
\end{table}
\begin{table}
\caption{van der Waals $C_{10}$ coefficients in $10^7$ a.u. for
homonuclear dimers. $C_{10}^v$ values include only
valence contributions. }
\label{C10hom}
\begin{ruledtabular}
\begin{tabular}{lccccc}
 & \multicolumn{1}{c}{Li}  & \multicolumn{1}{c}{Na}
 & \multicolumn{1}{c}{K }  & \multicolumn{1}{c}{Rb}
 & \multicolumn{1}{c}{Cs}  \\
\hline
{$C_{10}^v$}                 & 0.734     & 1.12    & 5.18     & 7.55       & 14.7      \\
{Final}                      & 0.735     & 1.13    & 5.37     & 7.96       & 15.9  \\
                     \multicolumn{6}{c}{Other works} \\
\protect\citet{PatTan97}     & 0.7289    & 1.068   & 4.789    & 6.833      & 13.58      \\
\protect\citet{YanBabDal96}  & 0.73721(1)&         &          &            &            \\
\protect\citet{MarSadDal94}  & 0.7365    & 1.107   & 5.248    & 7.665      & 15.20      \\
\protect\citet{SpeLorMey93}  & 0.7306    & 1.113   & 5.431    &            &            \\
\protect\citet{MaeKut79}     & 0.6901    & 1.036   & 4.522    & 6.836      & 13.01      \\
\end{tabular}
\end{ruledtabular}
\end{table}
\begin{table}
\caption{van der Waals $C_8$ coefficients in $10^5$ a.u. for
heteronuclear dimers. $C_8^v$ values include only
valence contributions.}
\label{C8het}
\begin{ruledtabular}
\begin{tabular}{lcccccccccc}
 & \multicolumn{1}{c}{Li-Na}  & \multicolumn{1}{c}{Li-K} & \multicolumn{1}{c}{Li-Rb}
 & \multicolumn{1}{c}{Li-Cs}  & \multicolumn{1}{c}{Na-K} & \multicolumn{1}{c}{Na-Rb}
 & \multicolumn{1}{c}{Na-Cs}  & \multicolumn{1}{c}{K-Rb} & \multicolumn{1}{c}{K-Cs}
 & \multicolumn{1}{c}{Rb-Cs}  \\
\hline
{$C_8^v$}&0.982    &1.91    &2.26    &3.07    &2.18    &2.56    &3.43    &4.64    &6.13    &7.04 \\
{Final}  &0.988(11)&1.95(2) &2.34(3) &3.21(10)&2.24(3) &2.66(4) &3.62(12)&4.93(6) &6.62(19)&7.69(22)\\
                     \multicolumn{11}{c}{Other theoretical works} \\
\cite{PatTan97}    & 0.949 & 1.852&2.190&3.049&2.082   &2.444   &3.355   &4.531   & 6.162  & 7.111 \\
\cite{MarSadDal94} & 1.068 & 2.517&3.137&4.586&2.614   &3.250   &4.727   &5.123   & 7.547  & 8.120 \\
\cite{SpeLorMey93} & 0.978 & 1.911&      &    &2.174   &        &        &        &        &       \\
\end{tabular}
\end{ruledtabular}
\end{table}
\begin{table}
\caption{van der Waals $C_{10}$ coefficients in $10^7$ a.u. for
heteronuclear dimers. $C_{10}^v$ values include only
valence contributions.}
\label{C10het}
\begin{ruledtabular}
\begin{tabular}{lcccccccccc}
 & \multicolumn{1}{c}{Li-Na}  & \multicolumn{1}{c}{Li-K} & \multicolumn{1}{c}{Li-Rb}
 & \multicolumn{1}{c}{Li-Cs}  & \multicolumn{1}{c}{Na-K} & \multicolumn{1}{c}{Na-Rb}
 & \multicolumn{1}{c}{Na-Cs}  & \multicolumn{1}{c}{K-Rb} & \multicolumn{1}{c}{K-Cs}
 & \multicolumn{1}{c}{Rb-Cs}  \\
\hline
{$C_{10}^v$} &0.912    &2.07    &2.55    &3.73    &2.48    &3.04    &4.40    &6.3     &8.9     &10.6 \\
{Final}      &0.916    &2.10    &2.61    &3.84    &2.53    &3.13    &4.55    &6.6     &9.4     &11.3 \\
                     \multicolumn{11}{c}{Other theoretical works} \\
\cite{PatTan97}    & 0.8859&1.949& 2.356 &3.379   & 2.303  & 2.773  & 3.948  & 5.724  & 8.077  & 9.629 \\
\cite{MarSadDal94} & 0.982 &2.651& 3.413 &5.303   & 2.949  & 3.784  & 5.844  & 6.726  & 10.37  & 11.79 \\
\cite{SpeLorMey93} & 0.9058&2.139&       &        & 2.553  &        &        &        &        &       \\
\end{tabular}
\end{ruledtabular}
\end{table}


\end{document}